# The dielectric constant of $Na_{0.4}K_{0.6}Br$ and its large temperature variation

**V. Katsika-Tsigourakou*[1], P. Bekris [1]**

[1] *Section of Solid State Physics, Department of Physics, National and Kapodistrian University of Athens, Panepistimiopolis, 157 84 Zografos, Greece*

**PACS** 61.72.Ji, 62.20.Dc, 66.30.Fq, 66.30.-h, 61.72.Bb

Since polycrystals of alkali halides are highly useful as components in optical devices, a number of mixed crystals of NaBr and KBr have been prepared from melt by other workers. Among these crystals, it was reported that the polycrystal $Na_{0.4}K_{0.6}Br$ exhibits the strongest temperature variation of the dielectric constant. Here, we show quantitatively that this is due to the temperature variation of the ionic polarizability.

**1 Introduction** It has been suggested, long ago, that several physical properties (e.g., the dielectric constant [1], the compressibility [2], the conductivity [3]) of alkali halide mixed crystals can be determined in terms of the corresponding properties of the pure end members (for a relevant review see chapter 12 of Ref. [4]). Several experimental studies appeared [5-12] on single phased and multiphased mixed crystals of alkali halides as a result of the intensified interest when it was realized that polycrystals of alkali halides are highly useful as components in optical devices [13]. As a first attempt towards understanding these experimental results, a procedure has been suggested [14], which enables the estimation of the compressibility of the multiphased mixed crystals in terms of the elastic data of the end members alone (this will be summarized later in Section 3). The key point of that procedure proposed [14] was the consideration of the volume variation produced by the addition of a "foreign molecule" to a host crystal as a defect volume whose compressibility was calculated on the basis of a thermodynamic model (termed $cB\Omega$ model) that has been found of value for the calculation of the defect (formation and migration) parameters in a large variety of solids [15-21].

Here, we deal with the experimental study [22] that refers to multiphased mixed crystals of NaBr and KBr. In this study, Padma and Mahadevan [22] achieved the preparation of mixed crystals $Na_xK_{1-x}Br$ of NaBr and KBr from melt and their X-ray diffraction analysis indicated the existence of two phases in the mixed crystals. The growth of the following five systems was reported [22]: $Na_{0.2}K_{0.8}Br$, $Na_{0.4}K_{0.6}Br$, $Na_{0.5}K_{0.5}Br$, $Na_{0.6}K_{0.4}Br$, $Na_{0.8}K_{0.2}Br$ (the composition written is taken for crystallization, see their Table 1). Beyond the study of thermal parameters (Debye-Waller factor, mean square amplitude of vibration, Debye temperature and Debye frequency), Padma and Mahadevan performed electrical measurements in all these (polycrystalline) mixed systems in the temperature range $308°$ to $423°K$. Namely, they measured the dc and ac conductivity (labeled $\sigma_{dc}$ and $\sigma_{ac}$) as well as the (real part of the) dielectric constant ($\varepsilon$) and the dielectric loss factor ($\tan\delta$). They found that the values of $\sigma_{dc}$, $\varepsilon$, $\tan\delta$ and $\sigma_{ac}$ increase with increasing temperature. However, this increase is different for different mixed systems. The results obtained indicated that the bulk composition has *nonlinear* influences on the electrical parameters. (Such a behavior has been also observed earlier [23] for single crystals NaBr-KBr). This nonlinearity is found to increase with the increase in temperature in the $\sigma_{dc}$, $\varepsilon$, $\tan\delta$ and $\sigma_{ac}$ values. In particular, Padma and Mahadevan [22] plotted each of these parameters versus the composition x and a maximum was identified at x=0.4 (see their figures 5, 6, 7 and 8). This maximum was very pronounced at the highest temperature ($423°K$) studied. In other words, among the five mixed systems mentioned above, the second one, i.e., $Na_{0.4}K_{0.6}Br$ exhibited a maximum in the corresponding plots for each of the parameters

$\sigma_{dc}$, $\varepsilon$, $\tan\delta$ and $\sigma_{ac}$ versus x, which was very pronounced for the temperature of $423^oK$.

In this paper we solely focus on the explanation of the variation of $\varepsilon$ with temperature for which Padma and Mahadevan [22] offered the following qualitative remarks: This is generally attributed to the crystal expansion, the electronic and ionic polarizations and the presence of impurities and crystal defects. The variation at low temperatures is mainly due to the expansion and electronic and ionic polarizations. In the case of alkali halide crystals, the electronic polarizability has no role to play. The increase at higher temperatures is mainly attributed to the thermally generated charge carriers and impurity dipoles. So, the observed increase in dielectric constant with temperature is essentially due to the temperature variation of ionic polarizability.

Before proceeding, and concerning the aforementioned qualitative arguments of Padma and Mahadevan [22], we note the following. Analytical reagent grade samples of NaBr and KBr were the starting materials they used for the growth of crystals. As they noticed, the dominant impurities present in NaBr including iron (0.001%) and those present in KBr included the divalent cations (calcium 0.001% and magnesium 0.001%). No specific controls were provided to prevent these impurities from entering the crystals. Actually, early studies [24] have shown that in alkali halides the presence of the aforementioned divalent cations produce (for reasons of charge compensation) additional cation vacancies; at low temperatures, a portion of these vacancies –which depends on temperature- are "bound", i.e., they are attracted by the nearly divalent cations, thus forming electric dipoles (usually termed complexes [24]) that contribute to the real (and imaginary) part of the dielectric constant. The remaining "free" vacancies (i.e., located far away from the impurities), contribute to the dc conductivity of the material (this contribution raises the so called extrinsic region [25] of the conductivity plot $\ell n(\sigma T)$ vs $1/T$ ). A theoretical calculation of the extent to which this phenomenon contributes to the observed temperature dependence of the dielectric constant is tedious and, at the present stage, not possible, since it demands the knowledge of the "association" (and "dissociation") parameters that govern the population of the electric dipoles "divalent impurity – cation vacancy" at each temperature. Unfortunately, these parameters have not been reported either in Ref. [4] or in other publications. In view of this difficulty, we restrict ourselves here to the calculation of the contributions of the electronic and ionic polarizability to the temperature dependence of the dielectric constant. Along these lines, a theoretical model is presented in the next section.

**2 The model that accounts for the temperature dependence of the electronic and ionic polarizability**

Here, we extend an early model suggested in Refs. [26-28] for the pure alkali halides to the case of mixed systems.

Szigeti [29], within the frame of classical theory, proposed the following two relations that interconnect the low frequency ($\varepsilon$) and high frequency ($\varepsilon_\infty$) dielectric constant of (undoped) alkali halides with the transverse optical mode frequency $\omega_{T_O}$:

$$f = \frac{\varepsilon + 2}{\varepsilon_\infty + 2} \mu \omega_{T_O}^2 \qquad (1)$$

$$\varepsilon - \varepsilon_\infty = \frac{4\pi}{9\upsilon} \frac{(\varepsilon_\infty + 2)^2 e^{*2}}{\mu \omega_{T_O}^2} \qquad (2)$$

where $f$ is the short-range force constant, $\upsilon$ the volume per ion pair, $\mu$ the reduced mass and $e^*$ the so called Szigeti effective charge. If we denote $\alpha_+$ and $\alpha_-$ the electronic polarizabilities of the cation and anion respectively, the classical Lorentz – Lorentz relation reads:

$$\frac{\varepsilon_\infty - 1}{\varepsilon_\infty + 2} = \frac{4\pi}{3\upsilon}(\alpha_+ + \alpha_-) \qquad (3)$$

while the low frequency (static) dielectric constant $\varepsilon$ can be calculated from:

$$\frac{\varepsilon - 1}{\varepsilon + 2} = \frac{4\pi}{3\upsilon}(\alpha_+ + \alpha_- + \alpha_{ion}) \qquad (4)$$

where $\alpha_{ion}$ stands for the ionic polarizability given by

$$\alpha_{ion} = e^2/f \qquad (5)$$

It has been found, however, that Eq. (4) agrees with the experimental data only when $\alpha_{ion}$ takes the value:

$$\alpha_{ion} = e^{*2}/f \qquad (6)$$

Considering that the number of cations (or anions) per unit volume is equal to $4/\ell^3$, where $\ell$ denotes the lattice constant in a NaCl structure, and that $\omega_{T_O}$ can be approximately calculated by the relation $\omega_{T_O}^2 = const \times B\ell/\mu$, where $B$ stands for the bulk modulus, a combination of the aforementioned relations finally leads to:

$$\frac{\varepsilon - 1}{\varepsilon + 2} - \frac{16\pi}{3\ell^3}(\alpha_+ + \alpha_-) = const \frac{1}{B\ell^4} \qquad (7)$$

in which $\ell$ depends on the temperature through the relation $\ell^3 = \ell_0^3 \exp\int_0^T \beta dT$, where $\beta$ is the thermal volume expansion coefficient. Obviously, the term in the right hand side of Eq. (7) refers to the ionic polarizability, while the term $\frac{16\pi}{3\ell^3}(\alpha_+ + \alpha_-)$ to the electronic polarizability.

In order to overcome the difficulty of the unknown value of the "*const*" involved in the right hand side of Eq. (7), we first apply this equation to a temperature $T_O$ (e.g., at room temperature, R.T.) and then determine the value of $\varepsilon$ at any other temperature (after disregarding, to a first approximation, corrections due to the volume dependence of the Szigeti charge). This leads to:

$$\varepsilon = \frac{2B_O\ell_O^4(1-\varepsilon_O)+2AB_O\ell_O(\varepsilon_O+2)-(\varepsilon_O+2)B\ell(2A+\ell^3)}{B_O\ell_O^4(\varepsilon_O-1)-AB_O\ell_O(\varepsilon_O+2)-(\varepsilon_O+2)B\ell(\ell^3-A)} \quad (8)$$

where the quantity $A$, defined as $A \equiv \frac{16\pi}{3}(\alpha_+ + \alpha_-)$, is known from the measured value of $\varepsilon_\infty$ through Eq. (3). Since, however $\varepsilon_\infty$ has not been measured in Ref. [22] at various temperatures, we have to rely on the approximation that $A$ is temperature independent. Furthermore, note that in Eq. (8) the values of the quantities $\varepsilon$, $B$ and $\ell$ correspond to any desired temperature T, while the subscript "o" denotes the relevant values at $T = T_O$.

**3 Application of Eq. (8) to the mixed system $Na_{0.4}K_{0.6}Br$** Let us now apply Eq. (8) to the system $Na_{0.4}K_{0.6}Br$ which, among the five mixed systems studied, exhibits the strongest temperature variation, as discussed above.

We take as $T_O$ the lowest temperature of 308°K for which the dielectric constant measured by Padma and Mahadevan to be 38.7. At this temperature, they reported the value of the compressibility $\kappa_O = 8.389 \times 10^{-11} m^2/N$. The latter gives for the corresponding bulk modulus $B_O(=1/\kappa_O)$ the ue× $B_O$ =11.92 GPa, for the mixed system. In order to apply Eq. (8), to the highest temperature (423°K) at which the dielectric constant has been measured in Ref. [22], we need to know the corresponding values of $B$ and $\ell$, which unfortunately have not been measured by Padma and Mahadevan [22] at temperature higher than 308°K. Hence, we must rely on certain approximations to estimate them:

We start with the lattice constant $\ell$, which is calculated as follows: The molar volume of the mixed crystal is given by $V = xV_1 + (1-x)V_2$, where $V_1$ and $V_2$ stand for the corresponding volumes of NaBr and KBr. Hence $\ell^3 = x\ell_1^3 + (1-x)\ell_2^3$, where $\ell_1$, $\ell_2$ correspond to the lattice constants of the pure end members KBr and NaBr respectively. (Thereafter, the subscripts "1" and "2" at all the quantities used will refer to KBr and NaBr, respectively). Thus, considering the values $\ell_1 = 6.60 \times 10^{-10} m$ and $\ell_2 = 5.98 \times 10^{-10} m$ for $T_O$ =308°K we find $\ell_O = 6.39 \times 10^{-10} m$ for the mixed system. The same is repeated for the highest temperature $T$ =423°K, by considering the corresponding values of the thermal expansion coefficient $\beta = 1.43 \times 10^{-4} K^{-1}$ for KBr and $\beta = 1.46 \times 10^{-4} K^{-1}$ for NaBr (obtained from an interpolation of their corresponding values of $\beta$ at room temperature and their melting temperature [30]). Thus we find the value $\ell = 6.42 \times 10^{-10} m$ at $T$ =423°K for the mixed system.

We now turn to the value of $B$. We first consider for KBr the compressibility values $44.03 \times 10^{-11}$ and $6.8 \times 10^{-11} m^2/N$ reported by Vaid et al [30] for the melting temperature (1007°K) and room temperature (R.T.), respectively, and therefrom we find the corresponding values of the bulk modulus. Then, by considering that in the high temperature range the $B_1$-value decreases almost linearly upon increasing the temperature [4], we find –by linear interpolation- that the corresponding $B_1$-value at 423°K is around 9.74 GPa. We now proceed to the calculation of $B$ for the mixed system, by following the procedure developed in Ref. [14]. This, in general, can be summarized as follows: Let $\upsilon_1$ be the volume per "molecule" of the pure component (1) (usually assumed to be the major component in the aforementioned mixed system). Without losing generality, we assume that $\upsilon_1$ is smaller than the volume $\upsilon_2$ per "molecule" of the pure component (2). Obviously $V_1 = N\upsilon_1$ and $V_2 = N\upsilon_2$, where $N$ stands for Avogadro's number. We now define a "defect volume" $\upsilon^d$ as the increase of the volume $V_1$ if one "molecule" of type (1) is replaced by one "molecule" of type (2). Thus, the addition of one "molecule" of type (2) to a crystal containing $N$ "molecules" of type (1) will increase its volume by $\upsilon^d + \upsilon_1$, where $\upsilon^d$ is a defect volume (see also below). If $\upsilon^d$ is independent of composition, the volume $V_{N+n}$ of a crystal containing $N$ "molecules" of type (1) and $n$ "molecules" of type (2) should be equal to:

$$V_{N+n} = N\upsilon_1 + n(\upsilon^d + \upsilon_1) \qquad (9)$$

which upon differentiating with respect to pressure leads to:

$$\kappa V_{N+n} = [1+(n/N)]\kappa_1 V_1 + n\kappa^d \upsilon^d \qquad (10)$$

where $\kappa$, $\kappa_1$ and $\kappa^d$ denote the compressibility of the mixed crystal, the end member 1 and the volume $\upsilon^d$, respectively. In the hard-spheres model, the "defect volume" $\upsilon^d$ can approximately determined from:

$$\upsilon^d = (V_2 - V_1)/N \text{ or } \upsilon^d = \upsilon_2 - \upsilon_1 \qquad (11)$$

and the compressibility $\kappa^d$ of the volume $\upsilon^d$ is given by:

$$\kappa^d = (1/B) - (d^2B/dP^2)/[(dB/dP)_T - 1] \qquad (12)$$

when considering the so-called $cB\Omega$ model, which has been found to be successful for describing the parameters for the formation and migration of the defects in a variety of solids [15-21]. Thus, since $V_{N+n}$ can be directly computed from Eqs. (9) and (11), the compressibility $\kappa$ of the mixed system is calculated from Eq. (10)

Let us now apply the aforementioned procedure to the case of the mixed system, i.e., $(NaBr)_{0.4}(KBr)_{0.6}$ in which the end member (pure) crystal (1) with the higher composition is of course KBr. In order to calculate $\kappa^d$, from Eq. (12), we use the following values for KBr: $(dB_1/dP)_T$=5.38 [31] and $d^2B/dP^2$=-0.492 GPa$^{-1}$ –obtained from the relation [14] $B_1(d^2B_1/dP^2) = -(4/9)(n^B+3)$, where $n^B$ is the usual Born exponent [31, 32]- along with the aforementioned value of $B_1$=9.74GPa at T=423°K. By inserting these values into Eq. (12) we find $\kappa^d$=21.50 GPa$^{-1}$. Furthermore, by considering the $\upsilon_1$ and $\upsilon_2$ values of KBr and NaBr respectively, we find $\upsilon^d$ from Eq. (11) and $V_{N+n}$ from Eq. (9), at T=423°K and finally obtain from Eq. (10) the value of $B(=1/\kappa)$=10.96 GPa for the mixed system at that temperature.

By inserting into Eq. (8) the values of $\ell$=6.42×10$^{-10}$m and $B$=10.96 GPa derived in the previous paragraphs, we find that the value of the dielectric constant $\varepsilon$ is $\varepsilon$=71.45. This agrees nicely with the experimental value 69.3 reported in Ref. [22].

**4 Discussion** We first comment on the fact that at 423°K the calculated value differs slightly, i.e. only by 3% (a difference which is anyhow between the experimental error), from the experimental one. This is remarkable, especially if we take into account that the $\varepsilon$-value measured at the highest temperature (T=423°K) is almost by a factor of around two larger than the corresponding value measured at the lowest temperature T=308°K. The origin for the success of Eq. (8) to account for such a considerable temperature dependence of $\varepsilon$, could be better understood from an inspection of Eq. (7), which reveals the following: In general, since $\ell^3$ exceeds $\ell_0^3$ at the most by a few percent, what accounts for the large temperature of $\varepsilon$ is the temperature decrease of the bulk modulus upon increasing the temperature. This means that, as the temperature increases, the ionic polarizability exhibits a considerable increase, which reflects a large temperature increase of $\varepsilon$ as well. In other words, Eq. (7) quantifies the merit of the qualitative argument of Padma and Mahadevan mentioned in the Introduction that the increase in dielectric constant with temperature is essentially due to the temperature variation of the ionic polarizability.

We now turn to an alternative usefulness of Eq. (7), which is of practical importance. Let us now assume that the temperature remains constant, but we vary the (external) pressure P. Then Eq. (7) predicts that $\varepsilon$ should also change upon varying the pressure mainly due to the pressure variation of B (cf. in ionic crystals the quantity $(dB/dP)_T$ usually reaches considerable values, i.e. around 5 or larger [4]). This seems to explain, in principle, the observation of the so called "coseismic" signals [33] i.e., the electric signals generated upon the arrival of the seismic waves at measuring site. (cf. These electric signals are entirely different from the precursory electric signals that are detected days to weeks before an earthquake occurrence [34, 35]). This is so, because the arrival of seismic waves causes a time variation of pressure around the measuring electrodes which reflects –according to Eq. (7)- a time variation of the polarization, thus producing electric signals (in a manner qualitatively similar to the well known piezoelectric phenomenon). These signals are intensified when a considerable density of charged dislocations is also present [33].

.

**5 Conclusion** Among the five polycrystalline mixed systems of NaBr and KBr, one of them, i.e., $Na_{0.4}K_{0.6}Br$, exhibits the strongest temperature variation in the dielectric constant $\varepsilon$. In particular, the $\varepsilon$-value at 423°K is about twice that at 308°K. Here, we showed that this variation can be almost exclusively accounted for from the temperature increase of the ionic polarizability.